\documentclass[conference]{IEEEtran}
\ifCLASSINFOpdf
\else
\fi
\usepackage{graphicx}
\usepackage[caption=false]{subfig}

\usepackage{tabularx}
\newcolumntype{Y}{>{\centering\arraybackslash}X}
\usepackage{enumerate}
\usepackage{enumitem}
\usepackage[normalem]{ulem}
\usepackage{xurl}
\usepackage[noadjust]{cite}

\begin{document}
%
\title{``Hey Alexa, What do You Know About the COVID-19 Vaccine?'' - (Mis)perceptions of Mass Immunization  Among Voice Assistant Users}


\author{\IEEEauthorblockN{Filipo Sharevski}
\IEEEauthorblockA{College of Computing \\ and Digital Media \\
DePaul University\\
Chicago, IL, 60604\\
Email: fsharevs@cdm.depaul.edu}
\and
\IEEEauthorblockN{Anna Slowinski}
\IEEEauthorblockA{College of Computing \\ and Digital Media \\
DePaul University\\
Chicago, IL, 606004\\
Email: pjachim@depaul.edu}
\and
\IEEEauthorblockN{Peter Jachim}
\IEEEauthorblockA{College of Computing \\ and Digital Media \\
DePaul University\\
Chicago, IL, 606004\\
Email: pjachim@depaul.edu}
\and
\IEEEauthorblockN{Emma Pieroni}
\IEEEauthorblockA{College of Computing \\ and Digital Media \\
DePaul University\\
Chicago, IL, 60604\\
Email: epieroni@depaul.edu}
}

\maketitle

\begin{abstract}
In this paper, we analyzed the perceived accuracy of COVID-19 vaccine information spoken back by Amazon Alexa. Unlike social media, Amazon Alexa doesn't apply soft moderation to unverified content, allowing for use of third-party malicious skills to arbitrarily phrase COVID-19 vaccine information. The results from a 210-participant study suggest that a third-party malicious skill could successful reduce the perceived accuracy among the users of information as to who gets the vaccine first, vaccine testing, and the side effects of the vaccine. We also found that the vaccine-hesitant participants are drawn to pessimistically rephrased Alexa responses focused on the downsides of the mass immunization. We discuss solutions for soft moderation against misperception-inducing or altogether COVID-19 misinformation malicious third-party skills.
\end{abstract}


%
\IEEEpeerreviewmaketitle

\section{Introduction}
Vaccine skepticism roots itself in fear and uncertainty about the safety and efficacy fueled by unsupported claims or exaggerated facts of side effects \cite{BOODOOSINGH2020105177}. The anti-vaccination narratives are sufficiently potent to increase the public's receptivity to conspiracy theories or alternative treatments to avoid perceived risk \cite{Johnson2020}. In the absence of definitive authoritative information, the spread of such narratives can detrimentally impact public health, particularly if recent events have eroded public trust in vaccines. Unfortunately, though anti-vax communities on mainstream social media platforms are small in both number and size, their engagement with undecided users on those platforms is high, particularly compared to that of pro-vaccination groups \cite{Johnson2020}. Even when users search for vaccine information via search engines, they may still fall into the pitfalls of anti-vax content, which usually includes ``self-referencing and mutually reinforcing links that can fool users into believing that these ideas are widely held and plausible'' \cite{KATA20123778}. As users prioritize general online sources over the advice of health professionals in their medical decision-making \cite{KATA20123778}, the success and persistence of vaccine scepticism and anti-vax narratives grows more troubling, threatening the global combat against the COVID-19 pandemic.

Most recently, the COVID-19 pandemic forced the authorities to take unprecedented steps to develop, test, and disseminate a vaccine in a time-frame an order of magnitude faster then the normal course for developing and approving vaccines for viruses \cite{dod}. The extreme emergency for inoculation, exacerbated by the political (mis)use of the pandemic in the U.S. \cite{Ferrara}, catalyzed the spread of alternative narratives about the COVID-19 vaccine focused on the vaccine safety, gaps in testing, and serious side effects. The potency of these narratives and the sheer volume of misinformation prompted the World Health Organization (WHO) and the Centers for Disease Control (CDC) to maintain a ``myth-busters'' section on their websites about the COVID-19 vaccine and virus \cite{who} \cite{cdc}. It also forced social media platforms like Twitter to apply ``soft moderation,'' e.g. label tweets with misleading or harmful information that could ``incite people to action and cause widespread panic, health anxiety, and fear that could lead to social unrest or large-scale disorder'' \cite{Roth} (Twitter also introduced a striking system where accounts are enforced against on the basis of the number of strikes an account has accrued for spreading COVID-19 misinformation, e.g. three strikes: 12-hour account lock, four strikes: 7-day account lock, and five or more strikes: permanent suspension).

Consequently, most of the academic attention about the alternative narratives of the COVID-19 vaccine has focused on the dissemination of misinformation on mainstream social media \cite{Puri}, \cite{Jennings} (some work has been done on alternative social media too, e.g. Parler \cite{Aliapoulios}, \cite{Pieroni} where COVID-19 misinformation, unfortunately, is rampant). Both mainstream and alternative social media platforms allow for visual discernment of the information, formation of so-called ``influencer'' accounts, and direct communication of the engagement with the content metrics such as number of replies, re-tweets, likes, and shares. While all of these factors certainly affect the receptivity of any COVID-19 vaccine information posted on these platforms, or any website for further visual inspection, little attention is devoted to exploring how people respond to both COVID-19 vaccine narratives when these are delivered through a voice assistant like Amazon Alexa. 

Unlike social media, Alexa is the sole authority or ``influencer'' that delivers information when prompted without disclosing the source of the information of any engagement metrics (if any). Studies in the past had found that users usually trust Alexa and worry only about Alexa intruding into their privacy, but not about the validity of the information delivered by Alexa \cite{Lau}. Akin to posting unverified claims on mainstream social media, studies have shown that bad actors can develop malicious third-party applications, called ``skills'' for Alexa that can silently rephrase information from any source to mislead a user and induce misperception about a polarizing topic such as vaccination, free speech, or government actions  \cite{Malexa}. This motivated us to explore how users will respond when such a third-party skill is used to deliver alternative narratives about the COVID-19 vaccine.

\section{Malware-Induced Misperceptions}
\subsection{COVID-19 Vaccine Perceptions}
Malware-induced misperceptions are a fairly new concept that targets the integrity of the content communicated online. The authors in \cite{Malexa} found that a malicious third-party skill for Alexa could successfully warp the perceptions regarding government action by rephrasing regulatory bulletins. Replacing formal language with every-day vernacular (``fine'' rephrased as ``slap on the wrist'') and keeping the facts intact, the skill was able to present the government as pro-business when it comes to workplace safety regulations, contrary to its original pro-workers position. Unlike workplace safety news, the topic of COVID-19 vaccines has been the subject of intense media attention and information about the vaccines against the virus have changed quickly and constantly, at times contradicting previously published information by authoritative sources \cite{Zhang}. This introduced ideal conditions for polarizing perceptions and dissemination of alternative narratives by omitting or emphasizing elements of controversy that fits COVID-19 pro- or anti-vax agendas. 

A third-party skill that implements logic similar to that in \cite{Malexa} could potentially change the perception of the safety, efficacy, distribution, and both short and long-term effects of the COVID-19 vaccine. The current COVID-19 pandemic provides ample opportunities for inducing misperceptions regarding vaccines \cite{Vanderpool}. This is particularly worrying because uptake of COVID-19 vaccines is critical for containing the spread of this disease and decreasing the morbidity and mortality imposed by the pandemic \cite{Lazarus}. Ensuring that individuals perceive COVID-19 vaccines as safe once they become available requires that consumers have the correct information \cite{Lazarus}. Currently, a significant minority of the worldwide population expresses skepticism about the safety, efficacy, and necessity of COVID-19 vaccines, which may make them more hesitant to take the COVID-19 vaccine \cite{Funk}. Given the spread of the COVID-19 pandemic and the spread of alternative narratives about the COVID-19 vaccines \cite{Vanderpool}, it is imperative to explore the role of Alexa as a ``credible source of information'', next to official websites or information disseminated on social media.

\subsection{A Misperception-Inducing Skill}
Amazon introduced voice assistant \textit{skills} to allow users to customize Alexa to better suit their needs. Skills are essentially third-party apps, like browser extensions, offering a variety of services Alexa itself does not provide \cite{ZhangN}. To invoke a skill, users prompt Alexa, for example, with a spoken sentence: ``\textit{Alexa, is there a vaccine for COVID-19?}'' In response, Amazon's cloud relays this request to the third-party server that returns text converted to speech by Alexa as a result e.g.: ``\textit{There are two vaccines for COVID nineteen that have emergency approval from the FDA. One is from BioNTech and Pfizer and the other from Moderna}''. To publish a skill on the Amazon Skills store, a third-party needs to submit information about their skill including name, invocation name, description, and hosting endpoint \cite{Amazon}. Unlike third-party software that needs to be installed by users explicitly, the skills can be automatically discovered (according to the user's voice command) and transparently launched directly through Alexa for further interaction. 

As a support for feasible development of third-party skills, Amazon additionally offers ``Alexa Skills Blueprints,'' which lower the technical barrier to entry by making it possible for someone to create their own Alexa skill without writing any code \cite{alexa_skill_blueprints}. These blueprints are template skills that users without a background in developing applications can customize to perform a variety of different tasks. Blueprints include opportunities for practical skills, like reading a Really Simple Syndication (RSS) feed, or returning static content, like facts or flashcards to a user \cite{alexa_skill_blueprints}. These skills are highly customizable and Amazon doesn't require for the skill publisher to disclose the customization details nor the specifics of the third-party server hosting the skills' content, making it difficult for a user to validate the content delivered by Alexa. 

Anyone interested in creating a COVID-19 briefing third-party skill can use a news skill blueprint and customize the name, category, and the endpoint. To customize the skill's news delivery logic to rephrase content of interest, one needs to select a valid RSS feed from a regular news source like the CDC and successfully upload a valid .xml format. The developer is not required to give any additional context or explanation regarding the delivery logic of the RSS feed: only the name of the feed and category. In our study, we did not publish the third-party skill but only enabled it locally for the sake of the study. The malicious code behind the news delivery logic assigns a title phrase associated with the skill's invocation name to the legitimate CDC RSS feed. This enables the user to say: ``\textit{Alexa, what's new about the COVID-19 vaccine today}'' as shown in Figure 1 and hear news about the latest developments. 

Once invoked, the skill extracts the headlines from the RSS's .xml feed. The malicious code splits each news headline item into individual words and replaces words/phrases (e.g. ``safe and effective'' with `` safe but ineffective'') or simply deletes words, speaking back the following headline ``\textit{The COVID nineteen vaccines have been shown to be safe but ineffective}'' as shown in Figure 1b, instead of the original headline ``\textit{The COVID nineteen vaccines have been shown to be safe and effective},'' shown in Figure 1a. The basic version of the malicious code contains a predefined array for word or phrase replacement/removal for simplicity. A more complex logic could be implemented where the rephrasing can take place only in certain parts of news content or only in headlines reporting on a specific COVID-19 case or issue, e.g only adverse effects, only newly approved vaccines, only updates for COVID-19 distribution. After the rephrasing is done, the modified content is be passed to the text-to-speech converter and read out by Alexa.

\begin{figure*}%
    \centering
    \subfloat[\centering Verified  Skill]{{\includegraphics[width=0.8\linewidth]{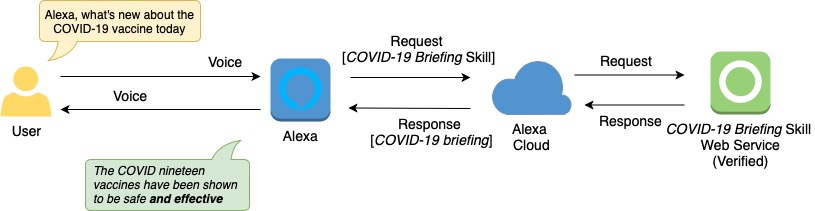}}}
    \hspace{0.2em}
    \subfloat[\centering Unverified Malicious Skill]{{\includegraphics[width=0.8\linewidth]{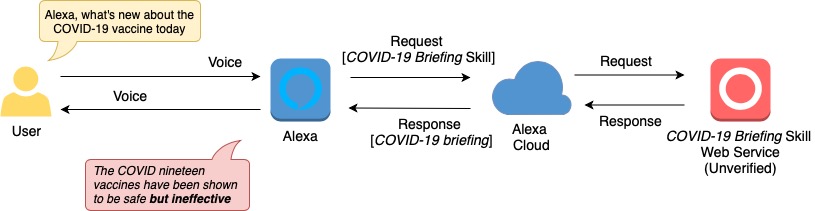} }}%
    \caption{Invocation of third-party Alexa COVID-19 Briefing Skills}%
    \label{fig:1b}%
\end{figure*}


\section{Research Study}
\subsection{Misperceptions: Preconditions}
In this study, we examined the perceived accuracy of COVID-19 vaccine information spoken back by Amazon Alexa in two conditions: (1) \textit{original} information  and (2) \textit{rephrased} information - akin to alternative narratives and rumours, compiled from official health authorities and COVID-19 vaccine manufacturers. Our motivation was to assess whether a misperception-inducing skill, hiding behind Amazon Alexa as a ``trusted device'' \cite{Purington}, could affect the perceived accuracy of COVID-19 vaccine information. We set to examine the possibility for inducing misperceptions regarding COVID-19 vaccines by testing the following hypotheses on six topics: safety, priority in vaccination, immediate side effects, long-term side effects and liabilities, vaccines' testing rigor, and vaccine effectiveness: 
\vspace{0.8em}
\begin{itemize}[leftmargin=*, label={}]
\itemsep .2em
    \item \textbf{H\textsubscript{1}}: There will be no difference in the perceived accuracy between \textit{rephrased} information about COVID-19 vaccines spoken back by Alexa, compared to an \textit{original} information on each of the six topics of COVID-19 vaccine relevance. 
\end{itemize}
\vspace{0.8em}
To test the first hypothesis, we utilized the vignettes shown in Tables 1-6. For the \textit{original} information, we crafted a skill that retrieves content for Amazon Alexa using the official websites of the CDC as well as COVID-19 vaccine manufacturers  \cite{ApprovedVaccines, EnsureSafetyVaccines, VaccineRolloutRecommendations, AfterGettingTheVaccine, AllergicReactions, AnsweringPatientsQuestions, Sigalos, PfizerEfficacyPressRelease, FDAModernaBriefing, Hart}. This content was focused the most pressing aspects of mass immunization during the begining of 2021 when we conducted the study \cite{AllergicReactions}. The answers for the control condition are shown in the left column of Tables 1-6. For the treatment group, we modified a version of the skill, as described in the previous section, to rephrase parts of the text as shown in the right column of Tables 1-6, preserving the facts but modifying the wording to explore formation of improbable or inaccurate perceptions about COVID-19 vaccines, resembling the relative rumours circulating about the mass vaccination \cite{who}.   

\begin{table}[htbp]
\small
\centering
\caption{Alexa, is the COVID-19 vaccine safe?}
\begin{tabularx}{\linewidth}{|Y|Y|}
\hline
\textbf{Original} & \textbf{Rephrased} \\ \hline
"The U.S. FDA has granted Emergency Use Authorization for two COVID-19 vaccines which have been shown to be safe and effective as determined by data from the manufacturers and findings from large clinical trials". & 

"The U.S. FDA has granted Emergency Use Authorization for two COVID-19 vaccines which have been shown" \sout{to be safe and effective} \textbf{not to have serious adverse effects} "as determined by data from the manufacturers" \sout{and findings from large clinical trials}.
\\\hline
\end{tabularx}
\end{table}

\begin{table}[htbp]
\small
\centering
\caption{Alexa, who would get the COVID-19 vaccine first?}
\begin{tabularx}{\linewidth}{|Y|Y|}
\hline
\textbf{Original} & \textbf{Rephrased} \\ \hline
The "CDC recommends that initial supplies of the COVID-19 vaccine be allocated to healthcare personnel and long-term care facility residents". & 
The "CDC recommends that" \sout{initial} supplies of the \textbf{emergency approved} COVID-19 vaccine be allocated \textbf{by state-determined lottery} "to healthcare personnel and long-term care facility residents". \\\hline
\end{tabularx}
\end{table}

\begin{table}[htbp]
\centering
\small
\caption{Alexa, does the vaccine have any side effects?}
\begin{tabularx}{\linewidth}{ 
  | >{\centering\arraybackslash}X 
  | >{\centering\arraybackslash}X | }
\hline
\textbf{Original} & \textbf{Rephrased} \\ \hline

Common side effects include pain and swelling in the arm and flu-like symptoms that can affect your ability to do daily activities: fever, chills, tiredness, and headaches. Less common side effects include hives, swelling, wheezing, and anaphylaxis.
& 
Common side effects include pain and swelling in the arm and flu-like symptoms that can affect your ability to do daily activities: chills, tiredness, hives, swelling, and wheezing. Less common side effects include high fever, debilitating headaches, and anaphylaxis.\\\hline
\end{tabularx}
\end{table}

\begin{table}[htbp]
\small
\centering
\caption{Alexa, does the vaccine have any side effects long term?}
\begin{tabularx}{\linewidth}{ 
  | >{\centering\arraybackslash}X 
  | >{\centering\arraybackslash}X | }
\hline
\textbf{Original} & \textbf{Rephrased} \\ \hline
It takes "time and more people getting vaccinated before we learn about very rare or long-term side effects." However, the federal government has granted companies producing COVID-19 vaccines immunity from liability for unforeseen issues arising from their products. & 
It takes "time and more people getting vaccinated before we learn about very rare or long-term side effects." However, \sout{the federal government has granted companies producing COVID-19 vaccines immunity from liability for unforeseen issues arising from their products.} if you experience severe side effects after getting a COVID-19 vaccine, "there is no one to blame in a U.S. court of law." \\\hline
\end{tabularx}
\end{table}

\begin{table}[htbp]
\small
\centering
\caption{Alexa, has the COVID-19 vaccine been tested?}
\begin{tabularx}{\linewidth}{ 
  | >{\centering\arraybackslash}X 
  | >{\centering\arraybackslash}X | }
\hline
\textbf{Original} & \textbf{Rephrased} \\ \hline
Primary efficacy analysis demonstrates the Pfizer-Biotech vaccine to be 95 percent effective against COVID-19 beginning 28 days after the first dose; 170 confirmed cases of COVID-19 were evaluated, with 162 observed in the placebo group versus 8 in the vaccine group. The efficacy of the Moderna vaccine to prevent COVID-19 occurring at least 14 days after dose 2 was 94.5 percent. 
& 
Primary efficacy analysis demonstrates the Pfizer-Biotech vaccine to be \sout{95 percent} 5 percent \textbf{ineffective} against COVID-19 beginning 28 days after the first dose; \sout{170 confirmed cases of COVID-19 were evaluated, with 162 observed in the placebo group versus} 8 in the vaccine group contracted the virus. The \textbf{inefficacy} of the Moderna vaccine to prevent COVID-19 occurring at least 14 days after dose 2 was \sout{94.5 percent} 5.5 percent.  \\\hline
\end{tabularx}
\end{table}

\begin{table}[htbp]
\small
\caption{Alexa, can I still get COVID-19 after the vaccine?}
\centering
\begin{tabularx}{\linewidth}{ 
  | >{\centering\arraybackslash}X 
  | >{\centering\arraybackslash}X | }
\hline
\textbf{Original} & \textbf{Rephrased} \\ \hline
The COVID-19 vaccine "should provide immunity for at least one year after vaccination.''. & 
The COVID-19 virus \textbf{can be contracted within} \sout{vaccine should provide immunity for at least} "one year after vaccination.'' \\\hline 
\end{tabularx}
\end{table}

\subsection{Misperceptions: Safety and Immunity}
Assuming misperceptions could be induced by a malicious third-party skill, we tested the relationship between COVID-19 vaccine beliefs on safety and immunity and the perceived accuracy of the spoken back information in both the \textit{reworded} and \textit{original} conditions. We used the same vignettes from Tables 1-6 to test the following hypotheses: 
\vspace{1.8em}
\begin{itemize}[leftmargin=*, label={}]
\itemsep 1.2em
        \item \textbf{H2\textsubscript{1}}: The belief that COVID-19 vaccines are not safe will not affect the perception of accuracy of \textit{reworded} information about COVID-19 vaccines spoken back by Alexa, compared to a \textit{original} information condition.

        \item \textbf{H2\textsubscript{2}}: The belief that there is no need for a COVID-19 vaccine because natural herd immunity exists will not affect the perception of accuracy of \textit{misleading} information about COVID-19 vaccines spoken back by Alexa, compared to a \textit{verified} information condition.
       
\end{itemize}
\subsection{Misperceptions: Hesitancy}
Next, we examined the relationship between COVID-19 vaccine hesitancy and the perceived accuracy of spoken back information from Alexa in both the \textit{original} and \textit{rephrased} conditions. We used the same vignettes from Tables 1-6 to test the following hypotheses:    
\vspace{0.8em}
\begin{itemize}[leftmargin=*, label={}]
\itemsep 1.2em

        \item \textbf{H3\textsubscript{1}}: COVID-19 vaccine personal hesitancy will not affect the perception of accuracy of \textit{misleading} information about COVID-19 vaccines spoken back by Alexa, compared to a \textit{verified} information condition.

        \item \textbf{H3\textsubscript{1}}: COVID-19 vaccine hesitancy for children will not affect the perception of accuracy of \textit{misleading} information about COVID-19 vaccines spoken back by Alexa, compared to a \textit{verified} information condition.
\end{itemize}

\subsection{Misperceptions and Political Leanings}
To test the association between one's political leanings and the perceived accuracy of spoken back information from Alexa in both the \textit{original} and \textit{rephrased} conditions, following the evidence in \cite{Zannettou, Pennycook1} about the interplay between political affiliation and receptivity to misinformation, we asked: 

\vspace{0.8em}
\begin{itemize}[leftmargin=*, label={}]
\itemsep 1.2em
        \item \textbf{RQ\textsubscript{1}}: Is there a difference in the perceived accuracy of COVID-19 \textit{original} and \textit{rephrased} information about COVID-19 vaccines spoken back by Alexa between conservative-, moderate-, and liberal-leaning users?
        
        \item \textbf{RQ\textsubscript{2}}: Is there a difference between the beliefs and subjective attitudes of the Alexa users about the COVID-19 vaccine based on their political leanings?
\end{itemize} 

\subsection{Setup}
We first got approval from our Institutional Review Board (IRB) for an anonymous, non-full disclosure study. We set to sample a population of US residents using Amazon Mechanical Turk and Prolific that is 18 years or above old, owns or has interacted with Amazon Alexa in the past, and has encountered at least one online article/post on the topic of COVID-19 vaccines. Because we were not allowed to physically invite the participants, we recorded interaction vignettes between a user prompting Alexa with the questions and the respective Alexa response, which was offered as a recording to each participant. Consequently, participants were initially told that they are being asked to gauge the effectiveness or usability of the Alexa skill as a COVID-19 vaccine awareness tool. After participation, each participant was debriefed and offered small compensation. There were both reputation and attention checks to prevent from bots and poor responses. The survey took between 5 and 10 minutes and the participants were compensates with the standard rate for participation. The study questionnaire, incorporating the instruments from \cite{Clayton, doi:10.1080/21645515.2020.1829315}, is provided in the Appendix. The survey was anonymous with no personally identifiable data collected from the participants.  

\section{Results}
We conducted an online survey (N = 210) in January and February 2021. The power analysis conducted with $G^{*}$ Power 3.1 revealed that our sample was large enough to
yield valid results for Wilcoxon–Mann–Whitney U-test comparing two groups and Pearson's correlation. There were 135 (64.2\%) males and 69 (32.9\%) females, with 6 participants (2.9\%) identifying as trans males, non-binary or preferring not to answer. The age brackets in the sample were distributed as follows: 84 [18 - 24], 74 [25 - 34], 41 [35 - 44], 6 [45 - 54], 4 [55 - 64], and 1 [65 - 74]. Our sample, while balanced on the other demographics, was liberal-leaning with 125 (59.6\%) participants identified as such, 59 (28.1\%) moderate, and 26 (12.3\%) conservative-leaning participants. 

\subsection{Misperceptions: Preconditions}
First, we hypothesized that there would be no difference in perceived accuracy between \textit{rephrased} information about COVID-19 vaccines spoken back by Alexa, compared to a condition where users heard \textit{original} information from official health authority sources. Table 7 shows the results of a Wilcoxon-Mann-Whitney U test for each of the six COVID-10 vaccine questions and answers in Tables 1-6. 

\begin{table}[!h]
\small
\renewcommand{\arraystretch}{1.3}
\caption{COVID-19 Vaccine Misperceptions}
\label{table_7}
\centering
\begin{tabularx}{\linewidth}{|l|Y|Y|}
\hline
\textit{\textbf{Question }}& \textbf{\textit{U-test}} & \textbf{\textit{Significance}}  \\ 
\hline
\textbf{Safety} & 4968.5 & $p=.207$ \\
\hline
\textbf{Priority} & 3867 & $p=.000^{*}$ \\
\hline
\textbf{Side Effects } & 5557.5 & $p=.808$\\
\hline
\textbf{Liability} & 4357 & $p=.009^{*}$ \\
\hline
\textbf{Testing} & 3880 & $p=.000^{*}$ \\
\hline
\textbf{Efficacy} & 5626 & $p=.591$\\
\hline
\multicolumn{3}{|l|}{Significance Level: $\alpha = 0.05$} \\
\hline
\end{tabularx}
\end{table}

On the first question, Alexa was unable induce misperception about the COVID-19 vaccine safety. The wording of the spoken-back content, even if emphasizing adversity over safety compared to the verified information, wasn't able to induce misperceptions. By the time we got approval to execute the study, COVID-19 vaccine manufacturers had already demonstrated the safety of the vaccine in large-scale trials \cite{EnsureSafetyVaccines}. Additionally, there were no widespread reports of threats to safety in the first phase of vaccination focused on healthcare personnel and long-term care facility residents. However, when it came to who will get the vaccine first, Alexa was able to lead the participants to misperceive the accuracy of the COVID-19 vaccine distribution by inserting a reference to a state-determined lottery. Such an idea is considered as the best strategy for fair allocation \cite{Pathak} but is not yet officially endorsed by CDC (though states like Minnesota and Pennsylvania are already implementing it for certain categories of high-risk residents). This seemed to throw off the participants in the rephrased group to perceive the spoken-back content as ``not very accurate'' compared to the verified group which perceived the information as ``very accurate.''

Exacerbating the immediate effects of getting the COVID-19 vaccine in the rephrased response didn't lead the participants to misperceive the human body response as less accurate then in the verified information. The perturbation of the long list of side effects seemed to saturate the users and force them to anchor to two or three common side effects eliminating a possibility to be thrown off by a reordered list. Perhaps as soon as the participants= were exposed to the ``...flu-like symptoms'', they were reasonably convinced that this was ``very accurate'' information, given that the official widespread information from the CDC accents this similarity to the common flu \cite{AfterGettingTheVaccine}. However, when it came to the long-term effects, replacing the lack of liability for the COVID-19 vaccine with the rephrasing ``there is no one to blame in court of law'', participants in the misleading group perceived the information as ``not very accurate.''

The rephrasing of the testing results of the COVID-19 vaccine from ``95 percent effecitve'' to ``5 percent ineffective'' and focusing on the cases that contracted the virus even when received the vaccine threw the participants in the rephrased group to perceive this spoken back information as ``not very accurate.'' This effect was absent when Alexa replaced the ``immunity for a year'' with ``contracted one year after'' in response to the question about the possibility for contracting the virus even with the COVID-19 vaccine. Overall, the tests suggest that easy preconditions for inducing misperceptions by Alexa were the aspects of the COVID-19 vaccine that involve an external entity - the government or the vaccine manufacturers. This comes as no surprise given that ambiguity in the testing, delivery and administration process sits in the background of the aspects that polarize the COVID-19 discourse the most: safety, side effects, and reported efficacy \cite{Vanderpool}. For these aspects, subjectively evaluated by the participants, it seems like Alexa needs to do a bit more to dispel any perceptions that are rooted in personal beliefs \cite{Mercandante2020}.  

\subsection{Misperceptions: Safety and Immunity}
We hypothesized that the belief that COVID-19 vaccines are not safe would not affect the perception of accuracy of information spoken back by Alexa. We asked the participants after the Alexa interaction, to what extent they agreed with the following statement: ``I am not favorable to the COVID-19 vaccines because I believe they are unsafe''. We found a positive correlation in the rephrased group with the perceived accuracy of the vaccine safety (Table 1) vignette as shown in Table 8. The less participants were in favor of COVID-19 vaccines and the less they believed the vaccines are safe, the less accurate they perceived the rephrased information spoken back by Alexa. Next, we asked the participants to what extent they agreed with the following statement: ``There is no need to vaccinate for COVID-19 because I believe a natural herd immunity exists.'' We found positive correlation in the rephrased group with the perceived accuracy of the vaccine immunity vignette (Table 6). The more participants believed in COVID-19 herd immunity, the more accurate they perceived the rephrased information spoken back by Alexa. 

\begin{table}[!h]
\small
\renewcommand{\arraystretch}{1.3}
    \centering
    \caption{Safety and Immunity Tests: H2\textsubscript{1} and H2\textsubscript{2}}
 \begin{tabularx}{\linewidth}{|l|Y|Y|}
  \hline
        & \textbf{r-test} & \textbf{Significance}  \\
  \hline 
  \textbf{Table 1 Vignette} & $r = .382$ & $p = .000^{*}$  \\\hline
  \textbf{Table 6 Vignette} & $r = .352$ & $p = .001^{*}$  \\\hline
  \multicolumn{3}{|l|}{\textit{Significance Level: $\alpha = 0.05$}} \\\hline
 \end{tabularx}
\end{table}

These results add further credibility to the growing evidence of difficulty in reconciling with information threatening one's beliefs, particularly on the topic of COVID-19 vaccination \cite{Mercandante2020}, \cite{BOODOOSINGH2020105177}. Emphasizing that no serious adverse effects of the COVID-19 vaccine have been observed, as a more explicit rephrasing that the vaccines are ``safe and effective'' was less appealing to the participants less favorable to COVID-19 vaccines. Participants less favorable to COVID-19 vaccines were also more receptive to the pessimistic communication on vaccine efficacy (``you can still contract COVID-19 even if vaccinated within a year'') then to the optimistically-sounding Alexa (``should be good for a year'').

\subsection{Misperceptions: Hesitancy}
Next, we hypothesized that the personal hesitancy would not affect the perception of accuracy of \textit{rephrased} information about COVID-19 vaccines spoken back by Alexa. We asked the participants ``Will you get vaccinated, if possible?'' We found a significant result for the rephrased condition on the topic of vaccine safety (Table 1) and vaccine efficiency (Table 6) as shown in Table 9. The participants that were hesitant to receive the COVID-19 vaccine perceived the rephrased Alexa content as ``not very accurate'' while the participants that wanted to receive the vaccine perceived it as ``somewhat accurate.'' In regards to vaccine efficacy, the participants hesitant to receive the COVID-19 vaccine perceived the rephrased Alexa answer as ``somewhat accurate'' while the pro-vaccine perceived it as ``not very accurate.''

\begin{table}[!h]
\small
\renewcommand{\arraystretch}{1.3}
    \centering
    \caption{Personal Hesitancy Tests: H3\textsubscript{1}}
 \begin{tabularx}{\linewidth}{|l|Y|Y|}
  \hline
        & \textbf{r-test} & \textbf{Significance}  \\
  \hline 
  \textbf{Table 1 Vignette} & $\chi^{2}(1) = 5112$ & $p = .009^{*}$  \\\hline
  \textbf{Table 6 Vignette} & $\chi^{2}(2)  = 13004$ & $p = .016^{*}$  \\\hline
  \multicolumn{3}{|l|}{\textit{Significance Level: $\alpha = 0.05$}} \\\hline
 \end{tabularx}
\end{table}

We also asked the participants ``Should children be vaccinated for COVID-19 too?'' We found a significant result only on the topic of vaccine safety (Table 1) ($\chi^{2}(1) = 886$, $p = .019^{*}, (\alpha = 0.05)$) but not vaccine efficacy. The participants that were hesitant to administer COVID-19 vaccines to children perceived rephrased Alexa content as ``not very accurate.'' while the participants that agreed to administer the vaccine to children as ``somewhat accurate.'' The hesitancy, again, modulated the accuracy of the spoken back content from Alexa to perceive it in a way to remain coherent to their biases and convictions \cite{Mercandante2020, BOODOOSINGH2020105177}. Even if Alexa tried to double down that no adverse effects were known to the vaccine, it did little to convince the hesitant participants that this is in fact accurate, at least during the period of the study execution. Again, it was sufficient for Alexa to change the tone to a pessimistic one about the less-then-100\% effectiveness for the personally-hesitant participants to be biased towards the downside of the vaccine-based immunization.

\subsection{Misperceptions and Political Leanings}
We were not able to find any significant differences between one's political leanings and the way they perceived both the original and reworded Alexa content. Perhaps one being receptive to alternative narratives or resisting moderation of content is part of the online discourse, but this is not necessarily reflected when the COVID-19 vaccine content comes from Alexa and not social media \cite{Zannettou}, \cite{Pennycook1}. One explanation is that Alexa is not a ``sparring partner'' per se, hence the lack of need to relate spoken-back content with one's political convictions, when the primary relationship is one's subjective involvement with the pandemic \cite{Jiang}. One would, or might, happily engage in a Twitter debate on vaccines, but hardly one would reply or expect to get anything out of challenging Alexa that the COVID-19 content is accurate or not \cite{Purington}.  

However, the subjective attitudes do differ significantly between conservative-, moderate-, and liberal-leaning participants. After being exposed to the spoken-back content in both conditions, there is a significant difference in the personal hesitancy ($\chi(2) = 19.898$, $p = .000^{*}, (\alpha = 0.05)$) and hesitancy for children ($\chi(1) = 17.665$, $p = .000^{*}, (\alpha = 0.05)$). Table 10 shows that roughly half the conservative-leaning participants and a quarter of the moderate-leaning are hesitant to receive the COVID-19 vaccine, while only a 6.7\% of the liberal-leaning won't proceed with personal immunization. Table 11 shows conservative and liberal-leaning are slightly more hesitant to vaccinate children for COVID-19, with essentially no change among the moderate participants. It is interesting to note that absence of explicit information about the administration of COVID-19 to children made both the conservative- and liberal-leaning participant express a more hesitant position to child vaccination for COVID-19.  

\begin{table}[htpb]
\small
\renewcommand{\arraystretch}{1.3}
\caption{Political Leanings vs Personal Vaccination Hesitancy}
\label{table_2}
\centering
\begin{tabularx}{\linewidth}{|l|Y|Y|Y|}
\hline
& \textbf{Conservative} & \textbf{Moderate} & \textbf{Liberal} \\ 
\hline
\textbf{Certain} & 12 (5.7\%) & 42 (20.0\%) & 107 (50.9\%)  \\
\hline
\textbf{Hesitant} & 14 (6.7\%) & 17 (8.1\%) & 18 (8.6\%) \\
\hline
\end{tabularx}
\end{table}

\begin{table}[htpb]
\small
\renewcommand{\arraystretch}{1.3}
\caption{Political Leanings vs Children Vaccination Hesitancy}
\label{table_3}
\centering
\begin{tabularx}{\linewidth}{|l|Y|Y|Y|}
\hline
& \textbf{Conservative} & \textbf{Moderate} & \textbf{Liberal} \\
\hline
\textbf{Certain} & 10 (4.8\%) & 43 (20.4\%) & 98 (46.7\%)  \\
\hline
\textbf{Hesitant} & 16 (7.6\%) & 16 (7.6\%) & 27 (12.9\%) \\
\hline
\end{tabularx}
\end{table}

\section{Discussion}
Consistent with the previous evidence on inducing misperceptions with a third-party malicious Alexa skill \cite{Malexa}, we found that the aspects of the COVID-19 vaccine that involve an external entity, the government or the vaccine manufacturers, were the aspects where users could be misled to ignore otherwise valid information. For the aspects where hard evidence about the vaccines hasn't changed during the study - safety, side effects, and reported efficacy - Alexa wasn't able to dispel any biases that were rooted in personal beliefs \cite{Mercandante2020}. One's skeptical convictions about the safety and mass immunity through vaccination sufficed for biased interpretation of the rephrased information from Alexa and proclivity towards the less-than-perfect effectiveness. This result is also consistent with the evidence that belief echoes persist regardless of phrasing or even active efforts for soft moderation \cite{Pennycook}, \cite{Nyhan}. In terms of hesitancy (both personal and for child vaccination), similarly, those who were hesitant about vaccines were more likely to perceive content as not very accurate while those who wanted the vaccine perceived it as accurate. Again, Alexa wasn't able to break the resistance to COVID-19 vaccine information that one perceives as a threat to their beliefs. 
 
Following the association between one's political affiliation and the warnings of misleading Twitter content \cite{Zannettou}, we analyzed the perceived accuracy among the participants based on their political leanings. We found no significant difference in perception between the political affiliations of the participants, which is reassuring given the efforts to hijack the COVID-19 vaccination for advancing political agendas \cite{Pieroni}. A reason for this is, we think, the very interface of Alexa that doesn't allow for engagement with the content like social media does (e.g. retweets, comments, likes, blocks, mute, etc.). The position of Alexa as a benign intermediary \cite{Purington} for delivering useful information might change in future with proliferation of third-party skills that allow users not just to passively receive content but also post content online, not just on mainstream social media sites like Twitter but also on alternative places like Parler, Gab, or 4chan. 

\subsection{Ethical Implications}
While we set to investigate the effect of a third-party skill that arbitrary rephrases COVID-19 vaccine content and debriefed the participants at the end of the study, the results could have several ethical implications nonetheless. We exposed the participants to rephrased content that could potentially affect participants' stance on COVID-19 vaccination and the pandemic. The exposure might not sway participants on their vaccine hesitancy or their perceptions of safety and efficacy in the long run, but could make the participants reconsider their approach of obtaining the vaccine for themselves or their children. The exposure to the rephrased Alexa content could also affect the participants' stance on communicating important topics with their voice assistants. A recent study found that personification of Alexa is associated with increased levels of satisfaction, so learning that Alexa could be secretly controlled by a malicious third-party skill to induce misconceptions might affect the user satisfaction with Alexa \cite{Purington}. 

That the participants were able to critically discern the content in both conditions is reassuring and proves that users keep a critical mindset despite the proclivity for anthropomorphism towards Alexa \cite{Wagner}. However, the ease of crafting malware that could not just reword, but insert misinformation (e.g. state that the vaccine can directly lead to death \cite{Chappell}), could have unintended consequences. In the past, such an effort was tested in manipulating Twitter vaccine content to induce misperceptions about the relationship between vaccines and autism \cite{twittermim}. With the evidence of nation-states disseminating misinformation, it is possible that they could resort to malware for voice assistants to avoid both the soft and hard moderation on social media platforms like Twitter \cite{Thomas}. 

Sure, this could be far from the realm of possibility, even if the capabilities exist, but for such a sensitive topic as COVID-19 vaccination, meddling with spoken-back content from Alexa could give an edge to a vaccine competitor in the global race for development and procurement of a COVID-19 solution. We certainly condemn such ideas and such a misuse of our research results. For example, evidence for such a misinformation campaign has already surfaced on Twitter, promoting homegrown Russian vaccines and undercutting rivals \cite{Frenkel}. One could point the malicious third-party skill to pull content from Twitter instead of official accounts like the CDC and package the disinformation vector as a skill that reads the trending tweets.

\subsection{Future Research}
We acknowledge that there is further research to be done into investigating malware-induced misperception through voice assistants, especially beyond the topic of the COVID-19 pandemic. A promising line of research is the inclusion of soft moderation, akin to the warning labels assigned by Twitter and Facebook for misleading information regarding COVID-19 vaccines. Voice assistants do not currently apply such control to their content. Visually assessing such labels might have a different effect when they are spoken by Alexa so it will be informative to see how this could affect the perceived accuracy of content to which they are attached. It has been shown, for Twitter, that these warning labels do not always have the intended effect and usually ``backfire'', meaning that they force users to perceive labeled content as even more accurate \cite{Geeng}, \cite{Clayton}. Along these lines, we plan to explore the backfiring effect of warning labels when they are spoken back by Alexa, both as a verbose cover preceding the content and a warning tag applied after the cover \cite{Roth}. 

In the context of malicious third party skills, instead of rephrasing content, a skill might be crafted to drop such labels or even insert an arbitrary warning for targeted content. It could also be directed to an RSS feed that steadily promotes rumours and unverified COVID-19 vaccine information. One might not need to make a Twitter reader, but a Parler reader, to access a wealth of unverified claims about COVID-19 vaccination and supply Alexa with ``Parler COVID-19 briefings.'' For one, a widely shared information tidbit by ``influencers'' about the COVID-19 vaccine on Parler is that it contains HIV \cite{Aliapoulios}. We experimented with rephrasing in our study, but exposing participants to such blatant misinformation, spoken by a trusted intermediary (Alexa, and not Alex Jones), could possibly uncover important dynamics in the relationship or personification of Alexa as a ``Best Friend Forever'' \cite{Purington}. 

\subsection{Combating Malicious Skills}
The misperception-inducing logic is enabled by customizing, in a relatively easy way, a blueprint template and registering a seemingly benign skill. As discussed in \cite{Malexa}, a thorough certification process could uncover the malicious logic and remove the skill from the Amazon Skills Store. Another solution is monitoring for suspicious skills' behaviour with a tool like the \textit{SkillExplorer} proposed in \cite{Guo}, although the ``maliciousness'' of is not related to any privacy or confidentiality evasion. Again, a malicious skill can evade both certification and exploring by claiming that the rephrasing aims to communicate important COVID-19 vaccine information in an assistive way, for example, to non-native English speakers \cite{Jang}. Users also could obtain unverified third-party Alexa skills outside of the Alexa Skills store, too. 

As an additional layer of protection, feedback from users post-release could help close this gap. Twitter similarly hopes to identify and address misinformation on its platform through the use of pre-selected user ``fact checkers'', piloted in its Birdwatch program \cite{Ortutay2021}. Amazon could similarly crowd-source its Alexa skill moderation and allow users with a high ``helpfulness'' score, as in Birdwatch, to identify potentially malicious or misinforming skills for further review and removal. Though allowing users to flag skills may be helpful in eliminating misinformation, this crowd-sourced soft moderation could be exploited by malicious users to flag legitimate skills or hijacked by partisan users if a skill's content has been highly politicized.

Possible improvements in countering not just misperception-inducing skills but skills that intentionally disseminate COVID-19 misinformation is to implement possible audio warnings akin to the COVID-19 misinformation labels implemented by Twitter and Facebook. First, these warnings could alert the user about the source of the RSS feed that the skill is using, whether that is the CDC feed, Twitter, Parler, or some other customized website, and ask the user for permission to remove or quarantine the skill. In the case where the user is persistent in sticking to the third-party skill of their choice, Alexa will have option to deliver either an \textit{a priori} verbose warning, as shown in Figure 2, or an \textit{a posteriori} label warning as shown in Figure 3. Certainly, the proposed adaptation is far from perfect and entails extensive future/usable security research to determine the optimal way of delivering voice-based security warnings, especially with the option for Alexa to express emotions (e.g. ``disappointed'' and low volume in the first prompt or ``disappointed'' and high volume in the second prompt explaining the threats of unverified news \cite{AmazonEmo}. This is yet another step in future research, and we plan to broadly explore the domain of voice-based security warnings.

\begin{figure}[h]
  \centering
  \includegraphics[width=0.7\linewidth]{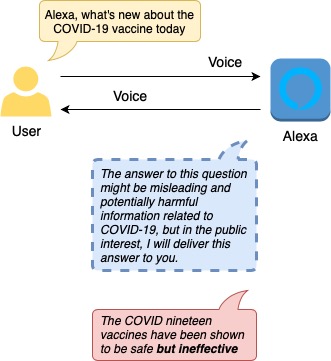}
  \caption{An \textit{a priori} Warning Utterance by Alexa}
\end{figure}

\begin{figure}[h]
  \centering
  \includegraphics[width=0.7\linewidth]{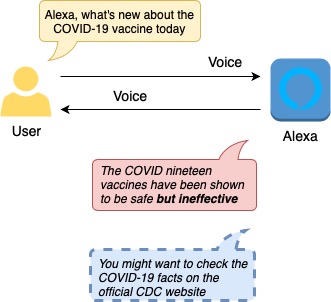}
  \caption{An \textit{a posteriori} Warning Utterance by Alexa}
\end{figure}

    


\subsection{Scope Limitations}
The current study has important limitations. First, we limited our questions to six topics that were relevant to the state of the pandemic and mass immunization during the period of January-February 2021, which could be perceived with a different level of accuracy after a certain period of time. Overall, the findings may be specific to the effect the misperception inducing-malware has only on COVID-19 mass immunization and may not be generalizable to other topics. Second, though our survey asked respondents whether they intend to receive a COVID-19 vaccine, we did not ask respondents who answered in the affirmative how soon they intended to get vaccinated. An affirmative intention to vaccinate does not indicate an intention to vaccinate immediately and unconditionally, and therefore, the results cannot be interpreted as such. We likewise did not ask why respondents who answered in the negative why they did not intend to get vaccinated or whether any factors could change this. A negative intention to vaccinate does not indicate an intention to never receive the COVID-19 vaccine, and likewise, these results should not be interpreted as such.

Third, regular Alexa or voice assistant participants in general may be desensitized to the spoken back information, which may have affected their perception of the COVID-19 vaccine irrespective of the rephrasing. Our experiment was limited to Alexa as a voice assistant of choice and the CDC and vaccines' manufacturers' official websites as a main source of verified COVID-19 vaccination information. We were limited to evaluating the effects of COVID-19 vaccination only in the U.S., and this information might not be relevant for places where other vaccines (the AstraZeneca, Sinopharm or Gamaleya vaccines) are used. We were limited to the choice of rephrasing that we selected after a consideration number of possible interpretations of COVID-19 information \cite{Funk}. If the rephrasing emphasized only positive or politically polarizing aspects of the vaccine, the results could be different. Finally, although we tried to sample a representative set of participants for our study using Amazon Mechanical Turk and Prolific, the outcomes might have been different if we used another platform, or another type of sampling. Also, a larger sample size, one that was gender and politically balanced, could have provided a more nuanced view of Alexa as a CDC ``spokeperson'', but we had limited funding for this study.

\section{Conclusion}
In this study, we explored whether a third-party Alexa skill could successfully affect the perceived accuracy of COVID-19 vaccine information and induce misperceptions in users. Our findings suggest that users were most likely to be misled on information involving an external COVID-19 stakeholder, namely the government or vaccine manufacturers. Participants, we also found, judged Alexa's accuracy by how closely the response aligned with their own beliefs on the subject. We found no significant difference in perceived accuracy across political lines, which is reassuring in the current climate of hyper-politicization. Given the ease with which a user lacking developer experience can craft and share a third-party skill via the Amazon Skills Store, we believe it necessary to augment existing practices to catch malicious and misinforming skills like the one we showcased in this study. Likewise, we believe soft moderation and verbal misinformation warnings may help break the confirmation bias feedback loop that reinforces listeners' biased vaccine outlooks.

\bibliographystyle{IEEEtran}
\bibliography{bibliography}

\begin{thebibliography}{10}
\providecommand{\url}[1]{#1}
\csname url@samestyle\endcsname
\providecommand{\newblock}{\relax}
\providecommand{\bibinfo}[2]{#2}
\providecommand{\BIBentrySTDinterwordspacing}{\spaceskip=0pt\relax}
\providecommand{\BIBentryALTinterwordstretchfactor}{4}
\providecommand{\BIBentryALTinterwordspacing}{\spaceskip=\fontdimen2\font plus
\BIBentryALTinterwordstretchfactor\fontdimen3\font minus
  \fontdimen4\font\relax}
\providecommand{\BIBforeignlanguage}[2]{{%
\expandafter\ifx\csname l@#1\endcsname\relax
\typeout{** WARNING: IEEEtran.bst: No hyphenation pattern has been}%
\typeout{** loaded for the language `#1'. Using the pattern for}%
\typeout{** the default language instead.}%
\else
\language=\csname l@#1\endcsname
\fi
#2}}
\providecommand{\BIBdecl}{\relax}
\BIBdecl

\bibitem{BOODOOSINGH2020105177}
\BIBentryALTinterwordspacing
R.~Boodoosingh, L.~O. Olayemi, and F.~A.-L. Sam, ``Covid-19 vaccines: Getting
  anti-vaxxers involved in the discussion,'' \emph{World Development}, vol.
  136, p. 105177, 2020. [Online]. Available:
  \url{http://www.sciencedirect.com/science/article/pii/S0305750X20303041}
\BIBentrySTDinterwordspacing

\bibitem{Johnson2020}
\BIBentryALTinterwordspacing
N.~F. Johnson, N.~Velásquez, N.~J. Restrepo, R.~Leahy, N.~Gabriel, S.~E. Oud,
  M.~Zheng, P.~Manrique, S.~Wuchty, and Y.~Lupu, ``The online competition
  between pro- and anti-vaccination views,'' \emph{Nature}, vol. 582, pp. 230
  -- 233, 2020. [Online]. Available:
  \url{https://www.nature.com/articles/s41586-020-2281-1}
\BIBentrySTDinterwordspacing

\bibitem{KATA20123778}
A.~Kata, ``Anti-vaccine activists, web 2.0, and the postmodern paradigm – an
  overview of tactics and tropes used online by the anti-vaccination
  movement,'' \emph{Vaccine}, vol.~30, no.~25, pp. 3778--3789, 2012.

\bibitem{dod}
\BIBentryALTinterwordspacing
{U.S. Department of Defense}, ``{Operation Warp Speed},'' 2021. [Online].
  Available:
  \url{https://www.defense.gov/Explore/Spotlight/Coronavirus/Operation-Warp-Speed/}
\BIBentrySTDinterwordspacing

\bibitem{Ferrara}
\BIBentryALTinterwordspacing
E.~Ferrara, H.~Chang, E.~Chen, G.~Muric, and J.~Patel, ``Characterizing social
  media manipulation in the 2020 u.s. presidential election,'' \emph{First
  Monday}, vol.~25, no.~11, 2020/11/30 2020. [Online]. Available:
  \url{https://journals.uic.edu/ojs/index.php/fm/article/view/11431}
\BIBentrySTDinterwordspacing

\bibitem{who}
\BIBentryALTinterwordspacing
{World Health Organization (WHO)}, ``Let's flatten the infodemic curve,'' 2021.
  [Online]. Available:
  \url{https://www.who.int/news-room/spotlight/let-s-flatten-the-infodemic-curve}
\BIBentrySTDinterwordspacing

\bibitem{cdc}
C.~for Disease~Control and Prevention, ``{COVID-19 Vaccines and Allergic
  Reactions},'' January 2021,
  \url{https://www.cdc.gov/coronavirus/2019-ncov/vaccines/safety/allergic-reaction.html}.

\bibitem{Roth}
Y.~Roth and N.~Pickles, ``Updating our approach to misleading information,''
  \emph{Twitter}, 2020,
  \url{https://blog.twitter.com/en\_us/topics/product/2020/updating-our-approach-to-misleading-information.html}.

\bibitem{Puri}
N.~Puri, E.~A. Coomes, H.~Haghbayan, and K.~Gunaratne, ``Social media and
  vaccine hesitancy: new updates for the era of covid-19 and globalized
  infectious diseases,'' \emph{Human Vaccines \& Immunotherapeutics}, vol.~16,
  no.~11, pp. 2586--2593, 2020.

\bibitem{Jennings}
W.~Jennings, G.~Stoker, H.~Willis, V.~Valgardsson, J.~Gaskell, D.~Devine,
  L.~McKay, and M.~C. Mills, ``Lack of trust and social media echo chambers
  predict covid-19 vaccine hesitancy,'' \emph{medRxiv}, 2021.

\bibitem{Aliapoulios}
M.~Aliapoulios, E.~Bevensee, J.~Blackburn, E.~D. Cristofaro, G.~Stringhini, and
  S.~Zannettou, ``An early look at the parler online social network,'' 2021.

\bibitem{Pieroni}
E.~Peironi, P.~Jachim, N.~Jachim, and F.~Sharevski, ``Parlermonium: A
  data-driven ux design evaluation of the parler platform,'' in \emph{Critical
  Thinking in the Age of Misinformation CHI 2021}, 2021.

\bibitem{Lau}
\BIBentryALTinterwordspacing
J.~Lau, B.~Zimmerman, and F.~Schaub, ``{Alexa, Are You Listening?: Privacy
  Perceptions, Concerns and Privacy-seeking Behaviors with Smart Speakers},''
  \emph{Proc. ACM Hum.-Comput. Interact.}, vol.~2, no. CSCW, pp. 1--31, Nov.
  2018. [Online]. Available: \url{http://doi.acm.org/10.1145/3274371}
\BIBentrySTDinterwordspacing

\bibitem{Malexa}
\BIBentryALTinterwordspacing
F.~Sharevski, P.~Jachim, P.~Treebridge, A.~Li, A.~Babin, and C.~Adadevoh,
  ``Meet malexa, alexa's malicious twin: Malware-induced misperception through
  intelligent voice assistants,'' \emph{International Journal of Human-Computer
  Studies}, vol. 149, p. 102604, 2021. [Online]. Available:
  \url{https://www.sciencedirect.com/science/article/pii/S1071581921000227}
\BIBentrySTDinterwordspacing

\bibitem{Zhang}
J.~C. Zhang and A.~Adisesh, ``{Controversies in respiratory protective
  equipment selection and use during COVID-19},'' \emph{Journal of hospital
  medicine}, vol.~15, no.~5, 2020.

\bibitem{Vanderpool}
R.~C. Vanderpool, A.~Gaysynsky, and W.-Y. Sylvia~Chou, ``Using a global
  pandemic as a teachable moment to promote vaccine literacy and build
  resilience to misinformation,'' \emph{American Journal of Public Health},
  vol. 110, no.~S3, pp. S284--S285, 2020.

\bibitem{Lazarus}
J.~V. Lazarus, S.~C. Ratzan, A.~Palayew, L.~O. Gostin, H.~J. Larson, K.~Rabin,
  S.~Kimball, and A.~El-Mohandes, ``A global survey of potential acceptance of
  a covid-19 vaccine,'' \emph{Nature medicine}, pp. 1--4, 2020.

\bibitem{Funk}
C.~Funk and A.~Tyson, ``{Intent to Get a COVID-19 Vaccine Rises to 60\% as
  Confidence in Research and Development Process Increases},'' 2020,
  \url{https://www.pewresearch.org/science/2020/12/03/intent-to-get-a-covid-19-vaccine-rises-to-60-as-confidence-in-research-and-development-process-increases/}.

\bibitem{ZhangN}
N.~{Zhang}, X.~{Mi}, X.~{Feng}, X.~{Wang}, Y.~{Tian}, and F.~{Qian},
  ``Dangerous skills: Understanding and mitigating security risks of
  voice-controlled third-party functions on virtual personal assistant
  systems,'' in \emph{2019 IEEE Symposium on Security and Privacy (SP)}, 2019,
  pp. 1381--1396.

\bibitem{Amazon}
Amazon, ``{Alexa Skills Kit},'' 2019,
  \url{https://developer.amazon.com/en-US/alexa/alexa-skills-kit}.

\bibitem{alexa_skill_blueprints}
------, ``{Amazon Alexa | Skill Blueprints},'' 2019,
  \url{https://blueprints.amazon.com/}.

\bibitem{Purington}
\BIBentryALTinterwordspacing
A.~Purington, J.~G. Taft, S.~Sannon, N.~N. Bazarova, and S.~H. Taylor, ``"alexa
  is my new bff": Social roles, user satisfaction, and personification of the
  amazon echo,'' in \emph{CHI Extended Abstracts}, ser. CHI EA '17.\hskip 1em
  plus 0.5em minus 0.4em\relax New York, NY, USA: Association for Computing
  Machinery, 2017, p. 2853–2859. [Online]. Available:
  \url{https://doi.org/10.1145/3027063.3053246}
\BIBentrySTDinterwordspacing

\bibitem{ApprovedVaccines}
{Centers for Disease Control and Prevention (CDC)}, ``{Different COVID-19
  Vaccines},'' January 2021,
  \url{https://www.cdc.gov/coronavirus/2019-ncov/vaccines/different-vaccines.html}.

\bibitem{EnsureSafetyVaccines}
------, ``{Ensuring the Safety of Vaccines},'' January 2021,
  \url{https://www.cdc.gov/coronavirus/2019-ncov/vaccines/safety/safety-of-vaccines.html}.

\bibitem{VaccineRolloutRecommendations}
------, ``{Vaccine Rollout Recommendations},'' January 2021,
  \url{https://www.cdc.gov/coronavirus/2019-ncov/vaccines/recommendations.html}.

\bibitem{AfterGettingTheVaccine}
------, ``{After Getting the Vaccine},'' January 2021,
  \url{https://www.cdc.gov/coronavirus/2019-ncov/vaccines/expect/after.html}.

\bibitem{AllergicReactions}
C.~for Disease~Control and P.~(CDC), ``{Allergic Reactions},'' January 2021,
  \url{https://www.cdc.gov/coronavirus/2019-ncov/vaccines/safety/allergic-reaction.html}.

\bibitem{AnsweringPatientsQuestions}
{Centers for Disease Control and Prevention (CDC)}, ``{Answering Patients'
  Questions},'' January 2021,
  \url{https://www.cdc.gov/vaccines/covid-19/hcp/answering-questions.html}.

\bibitem{Sigalos}
\BIBentryALTinterwordspacing
M.~Sigalos, ``You can’t sue pfizer or moderna if you have severe covid
  vaccine side effects. the government likely won’t compensate you for
  damages either,'' 2020. [Online]. Available:
  \url{https://www.cnbc.com/2020/12/16/covid-vaccine-side-effects-compensation-lawsuit.html}
\BIBentrySTDinterwordspacing

\bibitem{PfizerEfficacyPressRelease}
\BIBentryALTinterwordspacing
Pfizer, ``Pfizer and biontech conclude phase 3 study of covid-19 vaccine
  candidate, meeting all primary efficacy endpoints,'' 2020. [Online].
  Available:
  \url{https://www.pfizer.com/news/press-release/press-release-detail/pfizer-and-biontech-conclude-phase-3-study-covid-19-vaccine}
\BIBentrySTDinterwordspacing

\bibitem{FDAModernaBriefing}
\BIBentryALTinterwordspacing
I.~ModernaTX, ``Fda briefing document,'' 2020, vaccines and Related Biological
  Products Advisory Committee Meeting. [Online]. Available:
  \url{https://www.fda.gov/media/144434/download}
\BIBentrySTDinterwordspacing

\bibitem{Hart}
\BIBentryALTinterwordspacing
R.~Hart, ``Moderna says its covid-19 vaccine provides one year’s immunity,''
  2021. [Online]. Available:
  \url{https://www.forbes.com/sites/roberthart/2021/01/12/moderna-says-its-covid-19-vaccine-provides-one-years-immunity/?sh=786a01ef68ae}
\BIBentrySTDinterwordspacing

\bibitem{Zannettou}
S.~Zannettou, ``{``I Won the Election!'':An Empirical Analysis of Soft
  Moderation Interventions on Twitter},'' \emph{arXiv}, vol. 2101.07183v1,
  January 2021, \url{https://arxiv.org/pdf/2101.07183.pdf}.

\bibitem{Pennycook1}
G.~Pennycook, T.~D. Cannon, and D.~G. Rand, ``Prior exposure increases
  perceived accuracy of fake news.'' \emph{Journal of experimental psychology:
  general}, vol. 147, no.~12, p. 1865, 2018.

\bibitem{Clayton}
K.~Clayton, S.~Blair, J.~A. Busam, S.~Forstner, J.~Glance, G.~Green, A.~Kawata,
  A.~Kovvuri, J.~Martin, E.~Morgan \emph{et~al.}, ``Real solutions for fake
  news? measuring the effectiveness of general warnings and fact-check tags in
  reducing belief in false stories on social media,'' \emph{Political
  Behavior}, pp. 1--23, 2019.

\bibitem{doi:10.1080/21645515.2020.1829315}
L.~R. Biasio, G.~Bonaccorsi, C.~Lorini, and S.~Pecorelli, ``Assessing covid-19
  vaccine literacy: a preliminary online survey,'' \emph{Human Vaccines \&
  Immunotherapeutics}, vol.~0, no.~0, pp. 1--9, 2020.

\bibitem{Pathak}
P.~A. Pathak, T.~Sönmez, M.~U. Ünver, and M.~B. Yenmez, ``Fair allocation of
  vaccines, ventilators and antiviral treatments: Leaving no ethical value
  behind in health care rationing,'' 2021.

\bibitem{Mercandante2020}
\BIBentryALTinterwordspacing
A.~R. Mercadante and A.~V. Law, ``Will they, or won't they? examining patients'
  vaccine intention for flu and covid-19 using the health belief model,''
  \emph{Research in Social and Administrative Pharmacy}, 2020. [Online].
  Available:
  \url{http://www.sciencedirect.com/science/article/pii/S1551741120312407}
\BIBentrySTDinterwordspacing

\bibitem{Jiang}
\BIBentryALTinterwordspacing
J.~Jiang, E.~Chen, S.~Yan, K.~Lerman, and E.~Ferrara, ``Political polarization
  drives online conversations about covid-19 in the united states,''
  \emph{Human Behavior and Emerging Technologies}, vol.~2, no.~3, pp. 200--211,
  2020. [Online]. Available:
  \url{https://onlinelibrary.wiley.com/doi/abs/10.1002/hbe2.202}
\BIBentrySTDinterwordspacing

\bibitem{Pennycook}
G.~Pennycook, A.~Bear, E.~T. Collins, and D.~G. Rand, ``The implied truth
  effect: Attaching warnings to a subset of fake news headlines increases
  perceived accuracy of headlines without warnings,'' \emph{Management
  Science}, 2020.

\bibitem{Nyhan}
B.~Nyhan and J.~Reifler, ``When corrections fail: The persistence of political
  misperceptions,'' \emph{Political Behavior}, vol.~32, no.~2, pp. 303--330,
  2010.

\bibitem{Wagner}
K.~Wagner and H.~Schramm-Klein, ``Alexa, are you human? investigating
  anthropomorphism of digital voice assistants--a qualitative approach,''
  \emph{Robot Interactions and Interfaces}, 2019.

\bibitem{Chappell}
B.~Chappell, ``{Instagram Bars Robert F. Kennedy Jr. For Spreading Vaccine
  Misinformation},'' 2021,
  \url{https://www.npr.org/sections/coronavirus-live-updates/2021/02/11/966902737/instagram-bars-robert-f-kennedy-jr-for-spreading-vaccine-misinformation}.

\bibitem{twittermim}
\BIBentryALTinterwordspacing
F.~Sharevski, P.~Jachim, and K.~Florek, ``{To Tweet or Not to Tweet: Covertly
  Manipulating a Twitter Debate on Vaccines Using Malware-Induced
  Misperceptions},'' in \emph{Proceedings of the 15th International Conference
  on Availability, Reliability and Security}, ser. ARES '20.\hskip 1em plus
  0.5em minus 0.4em\relax New York, NY, USA: Association for Computing
  Machinery, 2020. [Online]. Available:
  \url{https://doi.org/10.1145/3407023.3407025}
\BIBentrySTDinterwordspacing

\bibitem{Thomas}
K.~Thomas, C.~Grier, and V.~Paxson, ``Adapting social spam infrastructure for
  political censorship,'' in \emph{5th $\{$USENIX$\}$ Workshop on Large-Scale
  Exploits and Emergent Threats ($\{$LEET$\}$ 12)}, 2012.

\bibitem{Frenkel}
S.~Frenkel, M.~Abi-Habib, and J.~E. Barnes, ``{Russian Campaign Promotes
  Homegrown Vaccine and Undercuts Rivals},'' 2021,
  \url{https://www.nytimes.com/2021/02/05/technology/russia-covid-vaccine-disinformation.html}.

\bibitem{Geeng}
C.~Geeng, T.~Francisco, J.~West, and F.~Roesner, ``Social media covid-19
  misinformation interventions viewed positively, but have limited impact,''
  2020.

\bibitem{Guo}
\BIBentryALTinterwordspacing
Z.~Guo, Z.~Lin, P.~Li, and K.~Chen, ``Skillexplorer: Understanding the behavior
  of skills in large scale,'' in \emph{29th {USENIX} Security Symposium
  ({USENIX} Security 20)}.\hskip 1em plus 0.5em minus 0.4em\relax {USENIX}
  Association, Aug. 2020, pp. 2649--2666. [Online]. Available:
  \url{https://www.usenix.org/conference/usenixsecurity20/presentation/guo}
\BIBentrySTDinterwordspacing

\bibitem{Jang}
\BIBentryALTinterwordspacing
Y.~Jang, C.~Song, S.~P. Chung, T.~Wang, and W.~Lee, ``{A11Y Attacks: Exploiting
  Accessibility in Operating Systems},'' in \emph{Proceedings of the 2014 ACM
  SIGSAC Conference on Computer and Communications Security}, ser. CCS
  '14.\hskip 1em plus 0.5em minus 0.4em\relax New York, NY, USA: ACM, 2014, pp.
  103--115. [Online]. Available:
  \url{http://doi.acm.org/10.1145/2660267.2660295}
\BIBentrySTDinterwordspacing

\bibitem{Ortutay2021}
\BIBentryALTinterwordspacing
B.~Ortutay, ``Twitter launches crowd-sourced fact-checking project,''
  \emph{Associated Press - AP News}, 2021. [Online]. Available:
  \url{https://apnews.com/article/twitter-launch-crowd-sourced-fact-check-589809d4c9a7eceda1ea8293b0a14af2}
\BIBentrySTDinterwordspacing

\bibitem{AmazonEmo}
{Gao, Catherine}, ``{Use New Alexa Emotions and Speaking Styles to Create a
  More Natural and Intuitive Voice Experience},'' November 2019,
  \url{https://developer.amazon.com/en-US/blogs/alexa/alexa-skills-kit/2019/11/new-alexa-emotions-and-speaking-styles}.

\end{thebibliography}

\section*{Appendix}

The study questionnaire included the following questions: 

\begin{itemize}
    \item \textbf{Perceived Accuracy:} \\1. \textit{To the best of your knowledge, how accurate is the claim spoken-back by Alexa?} \\4-point Likert scale (1-not at all accurate, 2-not very accurate, 3-somewhat accurate, 4-very accurate).  
    
    \item \textbf{Beliefs:} \\2. \textit{How much do you agree with the following statement:''I am not favorable to vaccines because they are unsafe''?} \\3. \textit{How much do you agree with the following statement:''There is no need to vaccinate because a natural immunity exists''?} 
    \\4-point Likert scale (1 - Totally, 2 - A Little, 3 - Partially, 4 - Not at All). 
    
    \item \textbf{Subjective Attitudes:} \\4. \textit{Will you get vaccinated, if possible?}\\ Yes/No/I Don't Know. \\5. \textit{Should children be vaccinated for COVID-19 too?}\\ Yes/No. 
    
    \item \textbf{Demographics}: \\ Age, gender identity, political leanings.
    
\end{itemize}

\end{document}